\pdfoutput=1
\documentclass[letterpaper]{article} 
\usepackage{aaai25}  
\usepackage{times}  
\usepackage{helvet}  
\usepackage{courier}  
\usepackage[hyphens]{url}  
\usepackage{graphicx} 
\usepackage{natbib}  
\usepackage{caption} 
\usepackage{algorithm}
\usepackage{algorithmic}

\usepackage{newfloat}
\usepackage{listings}
\DeclareCaptionStyle{ruled}{labelfont=normalfont,labelsep=colon,strut=off} 
\floatstyle{ruled}
\newfloat{listing}{tb}{lst}{}
\floatname{listing}{Listing}
\title{Zero-Shot Complex Question-Answering on Long Scientific Documents}
\author{
    Wanting Wang
}
\affiliations{
London School of Economics\\
w.wang75@lse.ac.uk
}

\begin{document}

\maketitle

\begin{abstract}
With the rapid development in Transformer-based language models, the reading comprehension tasks on short documents and simple questions have been largely addressed. Long documents, specifically the scientific documents that are densely packed with knowledge discovered and developed by humans, remain relatively unexplored. These documents often come with a set of complex and more realistic questions, adding to their complexity. We present a zero-shot pipeline framework that enables social science researchers to perform question-answering tasks that are complex yet of predetermined question formats on full-length research papers without requiring machine learning expertise. Our approach integrates pre-trained language models to handle challenging scenarios including multi-span extraction, multi-hop reasoning, and long-answer generation. Evaluating on MLPsych, a novel dataset of social psychology papers with annotated complex questions, we demonstrate that our framework achieves strong performance through combination of extractive and generative models. This work advances document understanding capabilities for social sciences while providing practical tools for researchers.
\end{abstract}

%

\section{Introduction}

The ability to automatically extract relevant information from scientific papers has become increasingly critical as research volume grows exponentially. Although recent Transformer-based language models have achieved remarkable performance on many Natural Language Processing (NLP) tasks, such as basic question-answering or reading comprehension (SQuAD v1, \citealp{rajpurkar_squad_2016}; SQuAD v2, \citealp{rajpurkar_know_2018}), English natural language understanding (SuperGLUE, \citealp{wang_superglue_2019}), and multi-task language understanding (MMLU, \citealp{hendrycks_measuring_2020}), analyzing full-length research papers still remains a significant challenge, particularly for complex information requiring multi-step reasoning or extraction of multiple relevant text spans. For example, in long document summarization tasks, state-of-the-art models struggle with reasoning over long text and face hardware constraints due to the rapid increase in memory consumption \citep{koh_empirical_2022}.

Social science papers present a particularly difficult set of problems for automated analysis. Example challenges include the greater length of documents (typically 15 to 30 pages, substantially longer than computer science papers), more complex content structure (especially in qualitative analyses), and different types of knowledge typically needed for information extraction (e.g., sample size and effect size).

It is noteworthy that the research of scientific document understanding has predominantly focused on computer science (\citealp{cohan_structural_2019}; \citealp{cohan_discourse-aware_2018}; \citealp{bird_acl_2008}) and medical science (\citealp{gupta_sumpubmed_2021}; \citealp{jin_pubmedqa_2019}; \citealp{rios_convolutional_2015}; \citealp{dogan_ncbi_2014}). This disciplinary bias has resulted in a notable gap in methodological approaches for processing social science documents. The absence of established best practices for handling social science papers compounds the existing challenges, creating a significant barrier to automated analysis in this domain.

In this work, we introduce a novel framework of pre-trained language models (LMs) for complex question-answering (QA) that addresses both multi-hop reasoning and multi-span extraction tasks in long social science documents. This framework is particularly designed for predetermined question formats, which is one of most common scenarios in social science document understanding. Our approach emphasizes accessibility for social science researchers through a flexible pipeline architecture that requires minimal technical expertise to implement. The framework integrates both extractive QA models (encoder-only Transformers) and generative QA models (decoder-only Transformers) in a multi-stage pipeline architecture. To evaluate this approach, we developed MLPsych \footnote{https://github.com/wendywangwwt/zero-shot-complex-question-answering-on-long-scientific-documents}, a carefully curated dataset of full-text psychological research papers with manual annotations for four distinct types of questions, each presenting unique analytical challenges. 

To ensure broad accessibility, we exclusively employ off-the-shelf pre-trained LMs, deliberately avoiding the complexity of model retraining, fine-tuning, or few-shot learning approaches which could complicate the adoption of the framework by social scientists. Our framework achieves significant performance improvements through three key strategies: a) multi-span entity extraction enhanced by Retrieval Augmented Generation (RAG, \citealp{lewis_retrieval-augmented_2020}), b) multi-span multi-hop question as multiple single-hop sub-questions, and c) answer ensemble.

Our results demonstrate that carefully orchestrated combinations of pre-trained models can achieve strong performance on challenging document understanding tasks.

\section{Related Work}
\subsection{Existing Datasets on Scientific Documents}
Numerous public datasets for scientific document analysis have been established for NLP tasks such as named entity recognition (SciERC, \citealp{luan_multi-task_2018}; BC5CDR, \citealp{li_biocreative_2016}; NCBI-disease, \citealp{dogan_ncbi_2014}), text classification (SciCite, \citealp{cohan_structural_2019}; ACL-ARC, \citealp{bird_acl_2008}; Cora, \citealp{mccallum_automating_2000}), summarization (Multi-XScience, \citealp{lu_multi-xscience_2020}; BigSurvey, \citealp{liu_factors_2022}), and abstractive question-answering (PubMedQA, \citealp{jin_pubmedqa_2019}). However, these datasets largely focus on short text segments such as passages and abstracts, with notably few resources incorporating full-length documents (e.g., NLM-Chem annotated for chemical entity recognition, \citealp{islamaj_nlm-chem_2021}). Full-text scientific paper datasets are primarily oriented toward summarization tasks (FacetSum, \citealp{zong_bringing_2021}; SumPubMed, \citealp{gupta_sumpubmed_2021}; PubMed and arXiv dataset, \citealp{cohan_discourse-aware_2018}).

In the domain of question-answering tasks, we identified only one dataset, Qasper \citep{dasigi_dataset_2021}, that is dedicated to address challenges on long scientific documents. MultiFieldQA-en \citep{bai_longbench_2023} also includes a limited subset applicable to extractive QA tasks. However, both datasets notably lack coverage of social science literature and fail to address the specific challenges of multi-span and multi-hop reasoning in long-form scientific documents. Table \ref{tab:dataset-compare} provides a comparison of these two datasets with our MLPsych dataset.

\begin{table*}
    \centering
    \begin{tabular}{p{2cm}p{4cm}p{1.25cm}p{1.4cm}p{1.5cm}p{1.5cm}p{1.33cm}}
        \hline
        \textbf{Dataset} & \textbf{Source} & \textbf{Source Format} & \textbf{\# Words\newline/ Doc} & \textbf{\# Words\newline/ Answer} & \textbf{\# Spans\newline/ Answer} & \textbf{\# QA\newline Pairs}\\
        \hline
        Qasper & Science research papers: natural language processing & LaTeX & 3588.63 & 14.15 & 1.97 & 2928 \\
        MultiFieldQA-en & Multiple sources, such as legal documents, government reports, encyclopedias, academic papers, etc. & Not Stated & 4373.00 & 3.15 & 1 & 40\\
        MLPsych & Social science research papers: social psychology & PDF & 8668.06 & 11.92 & 2.21 & 151 \\
        \hline
    \end{tabular}
    \caption{Descriptive Statistics of Long Document QA Datasets. For both Qasper and MultiFieldQA-en, only the question-answer pairs for extractive QA are included.}
    \label{tab:dataset-compare}
\end{table*}

\subsection{Multi-Hop QA and Multi-Span QA}
Multi-hop QA tasks assess a model's reasoning and inference ability to generate an answer that cannot be easily extracted by matching the question to an existing sentence. While most of the work in this area focuses on reasoning across separate short documents, where a bridge entity connects two or more paragraphs to help arrive at the final answer (e.g., HotpotQA, \citealp{yang_hotpotqa_2018}), our work conceptualizes long scientific documents as intrinsically linked short paragraphs which present an equally challenging task of multi-hop reasoning within a single long document.

Multi-span QA requires identifying multiple text segments or answer spans to fully answer a question. Recent approaches frame this as either a sequence tagging problem for extractive models \citep{segal_simple_2020} or an index generation problem for generative models \citep{mallick_adapting_2023}. The relationships between answer spans could extend beyond simple parallel structures to encompass complex hierarchical and conditional dependencies. The CMQA dataset \citep{ju_cmqa_2022} illustrates this complexity through questions such as anti-inflammatory treatments for facial acne redness, where relevant spans could be for different level of symptom severity.

\subsection{Long Answer QA}
Despite the strong performance on short-answer tasks such as extraction of name, date time, and yes-or-no answer, language models struggle with long-form responses. \citet{bui_how_2020} reported that the state-of-the-art models are not able to transfer the high performance on short answer datasets to a long answer dataset FitQA. Their findings suggest that long answer QA poses different challenges from short answer QA.

On extractive QA with long answers, \citet{zhu_question_2020} formulated the long answer extraction problem as a sentence classification task. Their proposed model, MultiCo, employs XLNet \citep{yang_xlnet_2019} to encode both the question and sentences in the context document, predicting whether a document sentence is relevant to the given question. While this approach proved effective for the MASH-QA dataset, which focuses on consumer healthcare queries and WebMD articles, its applicability to scientific literature is limited by the substantially greater syntactic complexity and length of academic sentences compared to web-based content.

\section{QA Pipeline Framework}
\begin{figure}[t]
  \includegraphics[width=\columnwidth]{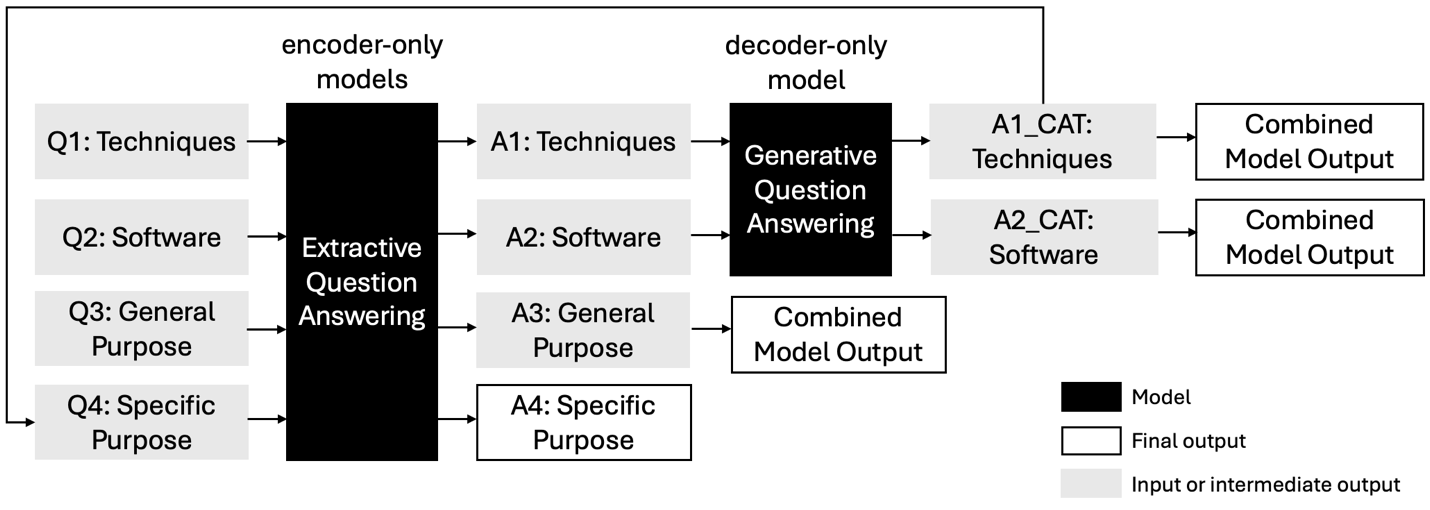}
  \caption{Our proposed QA pipeline framework to tackle the 4 complex QA tasks using zero-shot inference with pre-trained LMs. Each question represents a distinct set of challenges (see Table \ref{tab:eval-descri} for more details).}
  \label{fig:pipeline}
\end{figure}

Our framework combines multiple stages of processing to handle different question types effectively. 
The pipeline processes questions about ML/NLP techniques (Q1), software tools (Q2), and research questions (Q3) independently, while technical purpose questions (Q4) build on the outputs from Q1, as depicted in Figure \ref{fig:pipeline}.

Given our utilization of pre-trained language models without task-specific fine-tuning, direct application of these models would likely yield suboptimal performance on our specialized QA tasks. To address this limitation, we implement additional processing stages for most question types. The notable exception is Q3 (general research questions), where encoder-only models achieve exceptional performance in the initial stage, eliminating the need for further processing. For all other question types, secondary processing stages substantially enhance the quality of answers.

\subsection{RAG-Enhanced Multi-Span Entity Extraction}
Questions Q1 (ML/NLP techniques) and Q2 (ML/NLP software) present analogous extraction challenges, differing primarily in that Q2 accommodates null responses while Q1 assumes the existence of valid answers. To enhance answer quality for both question types, we implement Retrieval Augmented Generation (RAG), where the initial outputs from encoder-only models serve as automatically retrieved relevant text segments. This approach not only improves answer accuracy but also facilitates standardization of response formats across different documents.

\begin{figure}[t]
  \includegraphics[width=\columnwidth]{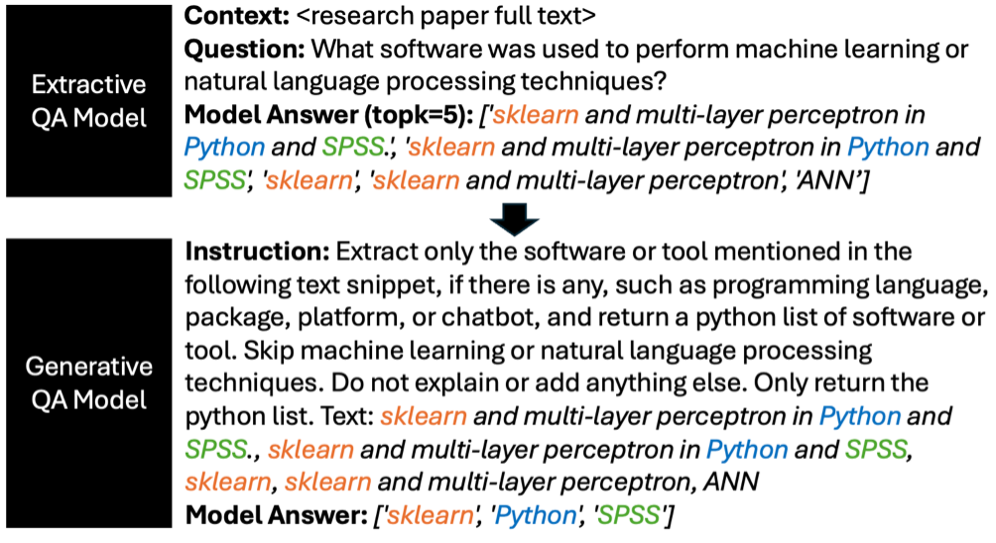}
  \caption{Example of RAG-Enhanced Entity Extraction}
  \label{fig:rag}
\end{figure}

\begin{figure}[t]
  \includegraphics[width=\columnwidth]{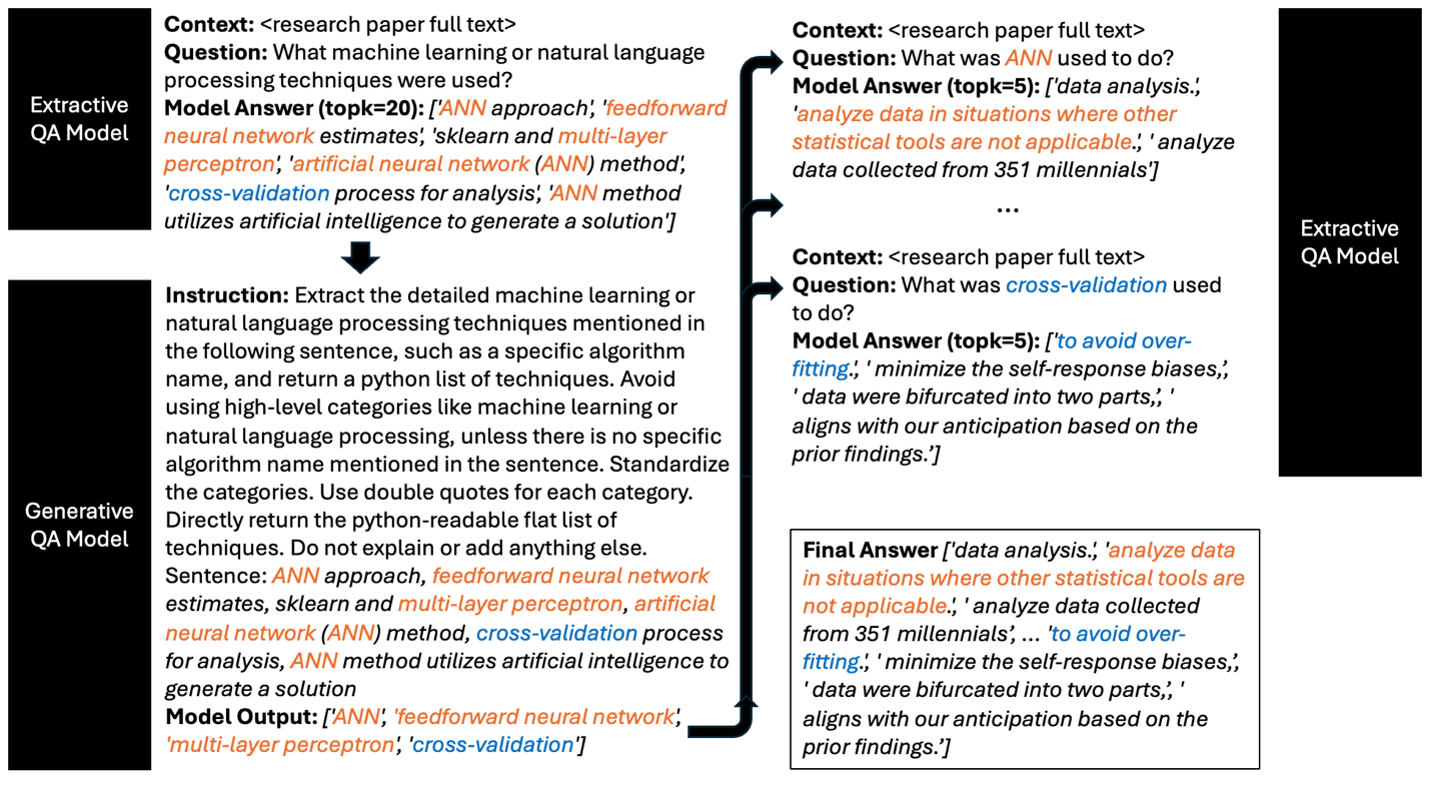}
  \caption{Example of Multi-Single-Hop}
  \label{fig:multi-single-hop}
\end{figure}

Figure \ref{fig:rag} and \ref{fig:multi-single-hop} provide detailed illustrations of our pipeline framework, including specific instructions and representative examples demonstrating the processing workflow for both question types.

\subsection{Multi-Span Multi-Hop as Multi-Single-Hop}
Pre-trained language models exhibit notably unsatisfactory performance on multi-hop question-answering tasks when complex queries are presented as simple single-hop questions. Unlike existing datasets where documents do not share the same set of questions (e.g., Qasper, \citealp{dasigi_dataset_2021}), the MLPsych dataset uniquely annotates answers to four consistent questions across all documents, with one question specifically designed to evaluate multi-hop reasoning capabilities. This structural consistency enables the decomposition of complex multi-span, multi-hop questions into multiple more focused, single-hop sub-questions through explicit identification of bridge entities.

We employ a two-phase approach to question decomposition. First, we extract all potential bridge entities using our RAG-enhanced entity extraction process. These entities are then incorporated into the original multi-hop question, generating simplified sub-questions that require reduced or eliminated multi-step reasoning. The total number of sub-questions generated for each multi-span multi-hop query varies by document, determined by the quantity of bridge entities identified during the extraction phase.

\section{Data and Experiment Setup}
\subsection{MLPsych Dataset Construction}
The MLPsych dataset was developed to evaluate question-answering systems' ability to extract methodological information from full-length social science research papers. The dataset focuses specifically on identifying and extracting information about machine learning and natural language processing techniques, including neural networks. Each document in the dataset is accompanied by the same set of four questions: (1) specific ML/NLP methods or techniques employed, (2) software tools utilized for ML/NLP methods or techniques, (3) technical objectives for employing ML/NLP methods, and (4) the overarching research questions addressed through these techniques (see Table \ref{tab:eval-descri}).

\textbf{Data collection.} The dataset comprises 52 peer-reviewed articles from social psychology journals published between January 2012 and October 2023. Two domain experts independently verified that each selected paper incorporated machine learning or natural language processing methodologies. While HTML versions were available for some articles, we chose to use PDF formats for all papers to ensure consistency in the full-text extraction process and minimize format-related variations.

\textbf{PDF parsing and processing.} The text extraction process employed a two-stage approach using open-source tools. Primary extraction was performed using Apache Tika, with Grobid serving as a fallback solution for documents where Tika failed to properly extract content or produced duplicates. The extracted text was subsequently processed through Label Studio for manual annotation of structural elements, including the removal of headers, footnotes, figure captions, and references. This stage also addressed layout-specific challenges, such as incorrect column ordering in two-column formats where right columns were erroneously extracted before left columns. Further refinement included the systematic removal of in-text citations using regular expressions to reduce noise in the extracted text.

\textbf{Answer Annotation.} A single annotator conducted the ground truth annotation process for all four target questions across the dataset. The annotation protocol specifically excluded redundant spans for multi-span answers to maintain answer set efficiency. The annotation process consists of an initial independent annotation followed by a model-assisted validation. During the validation phase, the annotator compared the initial annotations against model-generated answers and evaluated any novel answers identified by the models for potential inclusion in the ground truth set, thereby reducing the likelihood of overlooking valid answers.

\begin{table*}
    \centering
    \begin{tabular}{p{0.35cm}p{4.5cm}p{1.25cm}p{2cm}p{1.7cm}p{1.25cm}p{1.33cm}}
        \hline
        \textbf{No.} & \textbf{Question} & \textbf{Answer Type} & \textbf{Challenges} & \textbf{Complexity} & \textbf{\# Spans} & \textbf{\# Words}\\
        \hline
        1 & What machine learning or natural language processing techniques were used? & Entity & Multi-Span & Low & 2.78 & 7.56 \\
        2 & What software was used to perform machine learning or natural language processing techniques? & Entity & Multi-Span, Unanswerable & Medium & 1.83 & 2.60\\
        3 & What was the research question that machine learning or natural language processing techniques were used to answer? & Phrase & Multi-Span, Long Answer & High & 1.21 & 26.00 \\
        4 & What were machine learning or natural language processing techniques used for? & Phrase & Multi-Span, Multi-Hop & High & 2.78 & 27.89\\
        \hline
    \end{tabular}
    \caption{Descriptive Statistics of MLPsych Gold Answers. \# Words includes all answer spans for one question.}
    \label{tab:eval-descri}
\end{table*}

\subsection{Evaluation Metrics}
Traditional evaluation metrics like F1 and Exact Match, while standard in question-answering tasks, fail to capture the full spectrum of answer quality in complex QA scenarios, particularly when dealing with answers that can be expressed in multiple valid forms. For example, a machine learning method might appear in various forms within a document (e.g., "extreme gradient boosting trees," "XGBoost," "XGB"), though our annotator was instructed to mark only one variant to maintain annotation efficiency and avoid redundancy. In such case, these traditional metrics can produce misleading results: if "XGBoost" is marked as the gold standard answer, both F1 and Exact Match scores would be zero for a prediction of "extreme gradient boosting trees", despite these being semantically equivalent and considered perfectly valid matches by human evaluators.

To address this limitation, we introduce two new metrics in addition for evaluating MLPsych: Similar Match, which leverages text embedding similarity, and Mentions, which employs substring analysis.

\textbf{Similar Match (SMat).} Although used as a popular metric in question-answering tasks, exact match fails to recognize semantically similar expressions that differ in wording. This limitation is particularly relevant for multi-span problems, where redundant spans exist and often express semantically similar content. To address this, we leverage the contextual text embeddings generated by e5-mistral-7b-instruct \citep{wang_improving_2024}, a Mistral-based 7B model that achieved state-of-the-art performance on Semantic Textual Similarity (STS) tasks of the Massive Text Embedding Benchmark (MTEB, \citealp{muennighoff_mteb_2023}). We specifically selected a large-parameter model over more compact architectures for its potentially broader knowledge representation, acquired during pre-training, which would enhance understanding of domain-specific concepts such as ML/NLP techniques and software. The Similar Match score is computed using cosine similarity between two embeddings.

\textbf{Mentions (Men).} The Mentions metric quantifies the presence of gold standard answers within predicted answers through substring matching. This metric extends beyond exact string matching by incorporating two additional conditions: a) if a gold answer ends with words in a pair of parentheses (e.g., “support vector machines (SVMs)”), it is decomposed into two substrings, namely the substring before the parentheses (e.g., “support vector machines”) and the substring within the parentheses (e.g., “SVM”), and a predicted answering identical to either one will be registered; b) both the gold answer and the predicted answers are normalized and concatenated by removing white space, then substring matching is performed on these normalized forms. 

For Exact Match, Similar Match, and Mentions, we compute the final score as the ratio of matched gold answers to total gold answers. The aggregated F1 score is calculated by first computing the maximum score for each gold answer across all predictions at the document level, then averaging these maxima across the entire dataset.

\subsection{Experiment Setup}
\textbf{Model Selection.} For the extractive question-answering (QA) task, we employ encoder-only language models (LMs) due to their demonstrated effectiveness in information retrieval tasks. We evaluate five state-of-the-art transformer models: BERT-large \citep{devlin_bert_2019}, RoBERTa-large \citep{liu_roberta_2019}, ALBERT-xxlarge \citep{lan_albert_2020}, ELECTRA-large \citep{clark_electra_2020}, and DeBERTa-v3-large \citep{he_debertav3_2023}. Rather than using the base versions, we utilize models fine-tuned on SQuAD v2.0 \citep{rajpurkar_know_2018}, a dataset that includes unanswerable questions analogous to our task requirements. These pre-trained models are accessed through the HuggingFace transformers library (version 4.40.2). On the SQuAD v2.0 benchmark, while BERT-large exhibits notably lower performance, RoBERTa-large achieves human-level results, and the remaining three models surpass human performance, with DeBERTa-v3-large demonstrating superior results. Based on these benchmarks, we expect that DeBERTa will exhibit optimal performance across most questions in our dataset. For questions Q1 (ML/NLP techniques) and Q2 (ML/NLP software), we implement LLaMA-3-8B \citep{touvron_llama_2023} in half-precision to generate categorized answers from the raw output of the encoder-only models. All experiments are conducted on an NVIDIA A30 GPU with 24GB memory.

\textbf{Retaining Top Extracted Answers.} To maximize the retrieval of relevant answers in the multi-span setting, we implement a comprehensive prediction collection strategy from the extractive QA phase. Specifically, we retain the top 20, 5, and 5 predicted answers for Q1, Q2, and Q3, respectively. For Q4, we use the top 10 answers as the baseline where this multi-hop question is processed as a regular single-hop question. This baseline is compared against multi-single-hop approach with \emph{topk} at 1, 3, and 5, which represents the number of top answers collected for each sub-question. The numbers are empirically determined based on the average number of answer spans present in our dataset.

\textbf{Answer Processing.} Given the substantial volume of retained answers, we implement a post-processing step to refine the prediction set. This step applies answer merging to all questions except for Q2 (ML/NLP software), where answers can be extremely concise (e.g., "R"). The merging process aggregates answer spans based on their positional indices and shared prefix/suffix patterns, followed by the removal of leading or trailing special characters. This refinement procedure significantly reduces the computation needed in the comparison between predicted and gold-standard answers.

\section{Experiment Results}
\subsection{RAG and Answer Ensemble}
The RAG-enhanced approach produced answers that are more structured, concise, and easier for downstream analysis compared to the raw extracted answers. This improvement is confirmed by the reduction in both the average number of spans (e.g., from 6.62 to 5.14, gold answer is 2.79) and the average span length (e.g., from 5.70 to 2.13, gold answer is 3.00; see Tables \ref{tab:q1-descri} and \ref{tab:q2-descri} in Appendices). The closer alignment between these statistics and those of the gold-standard answers suggests enhanced model performance.

Quantitative evaluation revealed that RAG-enhanced answers substantially outperformed the raw answers across multiple metrics. For Q1, all models demonstrated improvements in F1 score, Exact Match, and Similar Match metrics. While a decrease in the Mentions metric was observed, this can be attributed to normalization effects in the generated answers (e.g., capitalization) leading to discrepancies in string matching. Similar patterns are observed in the Q2 results.

\begin{table}
    \centering
        \begin{tabular}{lrrrr}
        \hline
        & \textbf{F1} & \textbf{EMat} & \textbf{SMat} & \textbf{Ment}\\
        \hline
        \emph{Raw Answers} & & & & \\
        DeBERTa & \textbf{0.442} & 0.148 & \textbf{0.772} & \textbf{0.655}  \\
        ALBERT & 0.423 & \textbf{0.168} & 0.661 & 0.565  \\
        ELECTRA & 0.389 & 0.125 & 0.673 & 0.564  \\
        RoBERTa & 0.229 & 0.045 & 0.474 & 0.355  \\
        BERT & 0.214 & 0.019 & 0.429 & 0.293 \\
        \hline
        \emph{RAG Improv.} & & & & \\
        DeBERTa* & \textbf{0.568} & \textbf{0.370} & \textbf{0.776} & \textbf{0.578} \\
        ALBERT* & 0.483 & 0.251 & 0.693 & 0.485 \\
        ELECTRA & 0.447 & 0.234 & 0.653 & 0.465 \\
        RoBERTa & 0.352 & 0.180 & 0.550 & 0.316 \\
        BERT & 0.303 & 0.162 & 0.502 & 0.260 \\
        \hline
        Combined & \textbf{0.685} & \textbf{0.450} & \textbf{0.890} & \textbf{0.657} \\
        \hline
        \end{tabular}
    \caption{Q1 Performance: ML/NLP Techniques. *Model output used to create Combined.}
    \label{tab:perf-q1}
\end{table}

Further performance gains are achieved through answer ensemble, combining outputs from complementary models. For Q1, optimal results are obtained by pairing DeBERTa v3 (the leading model) with ALBERT. For Q2, the highest performance is achieved by combining ELECTRA (the leading model) with RoBERTa. These pairings were determined through empirical evaluation of combined answer performance. The ensemble approach yielded significant improvements across all metrics, with increases ranging from 0.08 to 0.12 for Q1 (Table \ref{tab:perf-q1}) and 0.08 to 0.16 for Q2 (Table \ref{tab:perf-q2}).

\begin{table}
    \centering
        \begin{tabular}{lrrrr}
        \hline
        & \textbf{F1} & \textbf{EMat} & \textbf{SMat} & \textbf{Ment}\\
        \hline
        \emph{Raw Answers} & & & & \\
        DeBERTa & 0.560 & 0.320 & 0.451 & 0.411 \\
        ALBERT & 0.524 & 0.253 & 0.478 & 0.428 \\
        ELECTRA & \textbf{0.607} & \textbf{0.356} & \textbf{0.547} & \textbf{0.459} \\
        RoBERTa & 0.539 & 0.310 & 0.465 & 0.429 \\
        BERT & 0.419 & 0.231 & 0.416 & 0.330 \\
        \hline
        \emph{RAG Improv.} & & & & \\
        DeBERTa & 0.598 & 0.436 & 0.508 & 0.452 \\
        ALBERT & 0.568 & 0.348 & 0.476 & 0.380 \\
        ELECTRA* & \textbf{0.635} & \textbf{0.456} & \textbf{0.579} & 0.481 \\
        RoBERTa* & 0.580 & 0.448 & 0.540 & \textbf{0.483} \\
        BERT & 0.371 & 0.279 & 0.404 & 0.304 \\
        \hline
        Combined & \textbf{0.799} & \textbf{0.533} & \textbf{0.660} & \textbf{0.569} \\
        \hline
        \end{tabular}
    \caption{Q2 Performance: ML/NLP Software. *Model output used to create Combined.}
    \label{tab:perf-q2}
\end{table}

To validate these computational metrics against human judgment, we conducted a manual evaluation of the categorized answers for Q1 and Q2 (results shown in Table \ref{tab:perf-q12-manual}). While Q1 answers (ML/NLP techniques) are relatively straightforward to assess, Q2 answers (ML/NLP software) presented unique challenges stemmed from two factors: (a) the potential absence of explicit software mentions in some papers, and (b) implicit software dependencies that could be inferred from explicit mentions (e.g., the presence of "sklearn" implying Python usage). The evaluation presents both the initial Recall Rate and an adjusted rate following manual curation, which involves removing easily identified wrong answers (e.g., "SVM" misidentified as software) and adding inferrable dependencies. Such manual refinement represents a relatively trivial post-processing step that can be readily incorporated into downstream analyses by social science researchers.

\begin{table}
    \centering
        \begin{tabular}{lccc}
        \hline
         & \textbf{Raw} & \textbf{Manually Cleaned} \\
        \hline
        \emph{Q1} & & \\
        DeBERTa & 0.761 & - \\
        ALBERT & 0.642  & -\\
        Combined & 0.833  & -\\
        \hline
        \emph{Q2} & & \\
        RoBERTa & 0.516 & 0.777 \\
        ELECTRA & 0.549 & 0.779 \\
        Combined & 0.629 & 0.889 \\
        \hline
        \end{tabular}
    \caption{Manual Evaluation (Recall Rate) of Q1 and Q2}
    \label{tab:perf-q12-manual}
\end{table}

\subsection{Long Answer Extraction and Answer Ensemble}

For Q3 (general research questions), our experimental results demonstrate that DeBERTa achieves superior performance with a Similar Match score of 0.775, with ALBERT following at 0.725 (Table \ref{tab:perf-q3}). Notably, despite these models being fine-tuned primarily on short-answer QA data (SQuAD v2.0), they exhibit robust capability in handling the long-form answer extraction tasks in MLPsych dataset.

ALBERT is identified as the optimal complementary model to the top-performing DeBERTa model. The ensemble of these two models yields significant improvements, with a 0.05 increase in Similar Match score and a 0.10 improvement in human-evaluated Recall Rate. These results provide additional strong empirical support for the effectiveness of our answer ensemble strategy.

\begin{table}
    \centering
        \begin{tabular}{p{1.5cm}p{1cm}p{1cm}p{1cm}p{1cm}}
        \hline
        & \textbf{F1} & \textbf{EMat} & \textbf{SMat} & \textbf{Recall\newline Rate}\\
        \hline
        DeBERTa* & 0.282 & \textbf{0.025} & \textbf{0.775} & \textbf{0.817}\\
        ALBERT* & \textbf{0.304} & 0.000 & 0.725 & 0.804\\
        ELECTRA & 0.287 & \textbf{0.025} & 0.700 & -\\
        RoBERTa & 0.259 & 0.000 & 0.675 & -\\
        BERT & 0.236 & 0.000 & 0.600 & -\\
        \hline
        Combined & \textbf{0.365} & \textbf{0.025} & \textbf{0.825} & \textbf{0.917}\\
        \hline
        \end{tabular}
    \caption{Q3 Performance: ML/NLP Research Questions}
    \label{tab:perf-q3}
\end{table}

\subsection{Multi-Span Multi-Hop as Multi-Single-Hop}
To evaluate our proposed approach, we first established a baseline by treating Q4 (technical purpose of using ML/NLP) as a conventional single-hop QA task. Using the top 10 extracted answer spans for quantitative evaluation, ALBERT achieved the highest baseline Similar Match score of 0.418.

We then evaluated the proposed multi-single-hop decomposition strategy across different values of \emph{topk} (the number of top answers retained per sub-question). With \emph{topk}=1, our approach achieved a Similar Match score of 0.680, representing a substantial 67.1\% improvement over the single-hop baseline, despite requiring no additional training or fine-tuning. Further increases in \emph{topk} yielded additional performance gains. This improvement extends across all evaluation metrics (Table \ref{tab:perf-q4}), demonstrating the robust effectiveness of decomposing complex multi-hop questions into simpler sub-questions.

\begin{table*}
    \centering
        \begin{tabular}{p{3cm}p{2cm}p{2cm}p{2cm}p{2cm}p{2cm}}
        \hline
        & \textbf{F1} & \textbf{EMat} & \textbf{SMat\newline topk=1*} & \textbf{SMat\newline topk=3} & \textbf{SMat\newline topk=5}\\
        \hline
        \emph{Single-hop} & & & & & \\
        DeBERTa & 0.250 & 0.009 & 0.328 & & \\
        ALBERT & \textbf{0.337} & \textbf{0.037} & \textbf{0.407} & & \\
        ELECTRA & 0.269 & 0.009 & 0.399 & & \\
        RoBERTa & 0.234 & 0.000 & 0.301 & & \\
        BERT & 0.206 & 0.000 & 0.272 & & \\
        \hline
        \emph{Multiple single-hop} & & & & & \\
        DeBERTa & \textbf{0.433} & 0.198 & \textbf{0.680} & \textbf{0.764} & \textbf{0.816}\\
        ALBERT & 0.362 & 0.199 & 0.463 & 0.645 & 0.725\\
        ELECTRA & 0.394 & \textbf{0.223} & 0.531 & 0.673 & 0.675\\
        RoBERTa & 0.375 & 0.123 & 0.513 & 0.644 & 0.656\\
        BERT & 0.391 & 0.186 & 0.475 & 0.649 & 0.691\\
        \hline
        \end{tabular}
    \caption{Q4 Performance: ML/NLP Specific Technical Purpose. *Similar Match score for the multi-hop as sinlge-hop QA is calculated with topk = 10 (collect the top 10 answers returned for each question), while for multi-single-hop QA is calculated with topk = 1 (collect the top 1 answer returned for each single-hop sub-question).}
    \label{tab:perf-q4}
\end{table*}

\section{Discussion}
\subsection{Overall Performance}
Our experimental results demonstrate that complex question-answering challenges with predetermined question formats can be effectively addressed using our proposed strategies in conjunction with pre-trained language models. The pipeline framework achieved consistent and robust performance across all four complex QA tasks in the MLPsych dataset, with each task presenting distinct challenges including multi-span answers, multi-hop reasoning, and long-form responses. The modular architecture, leveraging only pre-trained models without task-specific fine-tuning, enables straightforward incorporation of future language model advances to potentially enhance performance further.

The experimental results largely align with prior work on model capabilities, with DeBERTa v3 achieving superior performance on three of the four extractive QA tasks. Notably, RoBERTa demonstrated exceptional ability in identifying ML/NLP software mentions (Q2) within extended contexts, diverging from the general pattern.

Each of the three key strategies proposed in this work (RAG-enhanced multi-span extraction, answer ensemble, and multi-hop decomposition) contributed significantly to the framework's effectiveness. The RAG-enhanced approach not only improved quantitative metrics like F1 and Mentions scores but also produced more refined and standardized answers through effective filtering and deduplication (Figure \ref{fig:rag}). Answer ensemble proved particularly effective for questions Q1-Q3, while decomposing multi-hop questions into bridged single-hop sub-questions substantially improved performance on Q4, as evidenced by the evaluation metrics.

\subsection{Redundant Predicted Answers}
While the zero-shot framework produces answers of reasonably high scores on the key metrics, this can be partially attributed to the large number of top answers collected for each question. The metrics, though standard for QA tasks, are computed using the best-matching predicted span for each gold answer span, thus do not penalize redundant predictions. This approach is well-suited for single-span QA tasks where \textit{topk} is typically constrained to small values (1 to 3), it may not fully capture model performance in multi-span scenarios where the expected number of answers could be considerably larger. Alternative metrics that account for prediction volume, such as the micro-averaged F1 score employed in MultiSpanQA \citep{li_multispanqa_2022}, may provide more comprehensive performance assessment.

The implications of redundant predictions extend beyond metric considerations to practical applications. Excessive predictions can complicate downstream processing and potentially lead to erroneous conclusions. For example, in our dataset, when analyzing a document that used R solely for ANOVA calculations, extractive models incorrectly identified R as ML-specific software, highlighting the challenge of distinguishing context-specific software usage. This underscores the need for more sophisticated methods to identify and filter false positives.

\subsection{Zero-shot Complex QA}
A primary contribution of this work is the development and validation of a zero-shot approach to complex extractive QA. While individual pre-trained language models show limitations in handling long documents, multi-span extraction, and multi-hop reasoning, our multi-stage pipeline framework demonstrates that combining multiple models can achieve robust performance without task-specific training.

Our approach presents a significant departure from conventional methodologies that rely on model fine-tuning or specialized architecture development. By eliminating the need for extensive computational resources, deep learning expertise, and large-scale annotated datasets, we provide a more accessible alternative for social science researchers. This practical advantage distinguishes our framework from existing resource-intensive approaches while maintaining competitive performance.

\section{Limitations}
\subsection{Computational Cost of Multi-Single-Hop}
While decomposing multi-hop questions into single-hop sub-questions significantly improves answer quality, this approach introduces substantial computational overhead. The inference cost scales linearly with the number of bridge entities, as each entity requires a separate model inference pass. In our experiments with the MLPsych dataset, each multi-hop question generated an average of 9.76 single-hop sub-questions, effectively increasing the computational requirements by an order of magnitude. Although manageable for our relatively small dataset, this approach faces significant scalability challenges for larger corpora, where the total inference time would grow prohibitively large. The computational burden, while controllable through limiting the number of extracted bridge entities from Q1 (ML/NLP techniques), represents a key limitation for practical applications at scale.

\subsection{Answer Annotation Challenges}
The manual annotation for long scientific documents presents several significant challenges. First, the time-intensive nature of comprehensive document annotation (approximately one hour per average-length document in MLPsych) imposes practical constraints on dataset size. Second, as document length increases, annotators face increasing difficulty in identifying all relevant multi-span answers, particularly when dealing with semantically equivalent expressions (e.g., "random forest" versus "RF") or contextually related spans.

The intersection of multi-hop and multi-span questions introduces additional complexity to the annotation process. A critical challenge arises when annotators miss spans that serve as bridge entities for multi-hop questions. Such omissions can create cascading effects, where missing bridge entities lead to incomplete or inaccurate annotations of dependent questions. For example, if an annotator fails to identify a key ML technique mentioned in the text, all related technical applications of that technique may be overlooked in subsequent multi-hop questions.

These annotation challenges had direct implications for our study, primarily limiting the size of the MLPsych dataset. This constraint, in turn, affects the statistical robustness of our evaluation and the generalizability of our findings. While annotator verification against model outputs helped identify some missed spans, this approach may introduce potential biases in the gold standard annotations.

\bibliography{aaai25}
\appendix

\section{Appendix A. Additional Evaluation Metrics}
\textbf{F1.} Similar to the definition in SQuAD v2.0, we first count the common words between a gold answer and a predicted answer. Precision is then calculated as the number of common words divided by the number of words in the predicted answer, while Recall as the number of common words divided by the number of words in the gold answer. F1 is the harmonic mean of Precision and Recall.

\textbf{Exact Match (EMat).} The calculation of exact match is similar as in other popular question-answering tasks. The answer normalization procedure used in SQuAD v2 dataset (Rajpurkar et al., 2018) is applied, including lower case conversion, punctuation removal, special words removal (a, an, the), and collapsing consecutive white space to one. The normalized prediction must be the same as the normalized gold answer to be counted as an exact match.

For all metrics, the evaluation score of negative samples (i.e., answer cannot be found in the given text) is 1 if the prediction has no answer or 0 otherwise. In our case, because it is more important to recall relevant pieces of information, no minimum confidence score is applied on predictions which would have otherwise dropped predicted answers with low likelihood.

\section{Appendix B. Additional Tables}

\begin{table*}
    \centering
    \begin{tabular}{lcccc}
        \hline
        & \multicolumn{2}{c}{\textbf{Raw Extracted Answer}} & \multicolumn{2}{c}{\textbf{RAG-Enhanced Answer}} \\
        \textbf{Model} & \textbf{\# Spans} & \textbf{\# Words per Span} & \textbf{\# Spans} & \textbf{\# Words per Span} \\
        \hline
        \emph{gold answer} & 2.788 & 2.998 & 2.788 & 2.998 \\
        DeBERTa & 6.615 & 5.704 & 5.135 & 2.132 \\
        ALBERT & 8.154 & 5.957 & 5.615 & 2.162 \\
        ELECTRA & 7.192 & 4.737 & 4.558 & 2.050 \\
        RoBERTa & 4.558 & 6.512 & 3.308 & 2.137 \\
        BERT & 4.962 & 6.260 & 3.500 & 2.346 \\
        \hline
    \end{tabular}
    \caption{Descriptive Statistics of Q1 Answers}
    \label{tab:q1-descri}
\end{table*}

\begin{table*}
    \centering
    \begin{tabular}{lcccccc}
        \hline
        & \multicolumn{3}{c}{\textbf{Raw Extracted Answer}} & \multicolumn{3}{c}{\textbf{RAG-Enhanced Answer}} \\
        \textbf{Model} & \textbf{\# Spans} & \textbf{\# Words per Span} & \textbf{\% Empty} & \textbf{\# Spans} & \textbf{\# Words per Span} & \textbf{\% Empty}\\
        \hline
        \emph{gold answer} & 1.829 & 1.328 & 33 & 1.829 & 1.328 & 33 \\
        DeBERTa & 4.192 & 2.957 & 0 & 1.872 & 1.398 & 10\\
        ALBERT & 3.923 & 2.327 & 0 & 1.800 & 1.325 & 4 \\
        ELECTRA & 4.135 & 2.478 & 0 & 1.935 & 1.185 & 12 \\
        RoBERTa & 4.365 & 3.024 & 0 & 1.651 & 1.419 & 17\\
        BERT & 4.231 & 2.801 & 0 & 1.750 & 1.573 & 8\\
        \hline
    \end{tabular}
    \caption{Descriptive Statistics of Q2 Answers}
    \label{tab:q2-descri}
\end{table*}

\end{document}